# Characterization and remote sensing of biological particles using circular polarization

Lev Nagdimunov[1], Ludmilla Kolokolova[1*] and Daniel Mackowski[2]

[1]*University of Maryland, Department of Astronomy, MD, 20742, USA*

[2]*Auburn University, Mechanical Engineering Department, AL, 36849, USA*

* Corresponding author. tel. +1 (301) 405-1539
Email: *ludmilla@astro.umd.edu* (L. Kolokolova)

**ABSTRACT**

Biological molecules are characterized by an intrinsic asymmetry known as homochirality. The result is optical activity of biological materials and circular polarization in the light scattered by microorganisms, cells of living organisms, as well as molecules (e.g. amino acids) of biological origin. Lab measurements (Sparks et al. 2009a, b) have found that light scattered by certain biological systems, in particular photosynthetic organisms, is not only circular polarized but contains a characteristic spectral trend, showing a fast change and reversal of sign for circular polarization within absorption bands. Similar behavior can be expected for other biological and prebiological organics, especially amino acids. We begin our study by reproducing the laboratory measurements for photosynthetic organisms through modeling the biological material as aggregated structures and using the Multiple Sphere T-matrix (MSTM) code for light scattering calculations. We further study how the spectral effect described above depends on the porosity of the aggregates and the size and number of the constituent particles (monomers). We show that larger aggregates are characterized by larger values of circular polarization and discuss how light-scattering characteristics of individual monomers and electromagnetic interaction between them affect this result. We find that circular polarization typically peaks at medium (40-140°) scattering angles, and discuss recommendations for efficient remote observation of circular polarization from (pre)biological systems.

1. Introduction

An emerging interest in using circular polarization to detect the presence of biological materials remotely has developed over the last fifteen years. Stimulated in part by measurements of circular polarization in astronomical contexts and the astrobiological applications of such measurements, a possibility arises to employ circular polarization for remote sensing of objects that contain biological or pre-biological organics, including astronomical objects, terrestrial hydrosols, and aerosols as well as objects of bio-medical studies. In this paper we focus on particulate media, including microorganisms, dusts, and aerosols that contain either biological or pre-biological (e.g. amino acids and sugars) organics. For brevity, we call them "biological particles," "biological materials" or "biological objects."

Traditionally, astronomical circular polarization has been thought to be rarely present in the absence of high energetic phenomena. Small levels of circular polarization stemming from low energy processes have been technically difficult to measure. Nevertheless, it has now been reliably measured for planets, interstellar and interplanetary dust, and comets [1-4]. These papers also discuss the mechanisms that can be responsible for circular polarization: multiple scattering in asymmetric media, scattering by aligned particles, and scattering by intrinsically asymmetric particles. We explore properties of the latter effect in this study.

Scattering by intrinsically asymmetric particles is of particular interest in remote sensing of biological particles because their material is characterized by so called homochirality. By definition [5] a molecule is said to be homochiral if it is composed entirely of one type of chiral "building blocks", which exist in two non-superimposable mirror-symmetric forms. The vast majority of amino acids present in living tissue are "left-handed", and most sugars present in living materials use their "right-handed" mirror-symmetric version. This preference for a single handedness is unique to biological material, no non-biological objects have ever been found to possess it. Intrinsic asymmetry of biological molecules makes biological material optically active, i.e. able to produce circular polarization.

Homochirality would be an especially useful tool in a search for biological particles if it could be remotely detectable. While it has been suggested for as long as several decades that circular polarization resulting from light scattering by homochiral materials could be remotely detectable, any detection need to be separated from the other causes of circular polarization mentioned above. To assist with this problem, we studied how homochirality of biological particles affects the circular polarization in light scattered by them. This information would facilitate the identification of circular polarization measured from a remote object, and also potentially allow us to narrow down and describe the characteristics of the observed material.

Our work was primarily motivated by recent measurements of circular polarization for light scattered by organisms that contain photosynthetic pigments by Sparks et al. [6, 7]. Circular polarization produced by such pigments was found to be not only non-zero but also to vary distinctly with wavelength inside absorption bands, showing a rapid change in value and in some cases even a change in sign. An example of circular polarization measurements is shown in Fig. 1 for cyanobacteria WH8101 (adapted from [6]). Circular polarization shown in all figures of this paper is defined as the ratio of the forth Stokes parameter to the first one, i.e. V/I. Similar results were found for a variety of macroscopic vegetation as well as bacteria relying on photosynthesis for energy production. Control measurements of nonbiological chemical compounds and minerals showed either negligible circular polarization or noise-like variations of circular polarization not affected by a change in absorption [6, 7].

In this paper, we present the results of computer modeling for circular polarization produced by light scattering of chlorophyll *a*, and compare them to the lab measurements of polarization by cyanobacteria and macroscopic vegetation, in which chlorophyll *a* is both abundant and essential. We study chlorophyll due to the measurements discussed above and because its optical constants are better studied and can be easier found in the literature than those of other complex organics. However, we expect that certain amino acids should also show this effect as they demonstrate similar optical constant trends in their absorption bands [8-10].

We model biological material as aggregates of spherical particles, which provides the most realistic model (refer to section 3) of such particles that can be treated with available computational tools. To advance both the study of circular polarization in photosynthetic pigments as well as circular polarization of homochiral molecules in general, we performed a computational survey of the effect that properties of biological particles, and the aggregates we used to model them, have on circular polarization.

2. **Circular Polarization Induced by Optically Active Particles**

Circular polarization of light scattered by chiral molecules is caused by two material effects: optical rotation, sometimes also termed circular birefringence, and circular dichroism. Circular birefringence indicates that the material's real index of refraction is different for left and right circularly polarized light. Similarly, circular dichroism refers to different absorption for light of different handedness. Both effects, in the absence of a magnetic field, arise only in particles characterized by an intrinsic asymmetry.

There are four optical constants necessary to model circular polarization of light that is scattered by a chiral material. The first two are the real and imaginary parts of the complex index of refraction, describing refraction and absorption respectively. The other two parameters describe circular birefringence and circular dichroism. Mathematically, the complex index of refraction is $m = n+ik$, where n is different for left and right-handed polarized light in the presence of circular birefringence and k is different in the presence of circular dichroism; $i$ is the imaginary unit. We can create a parameter β, which describes the difference in the complex index of refraction between left and right-handed circularly polarized light: $β=m_L-m_R=β_R+iβ_I$, where $β_R$ represents circular birefringence, $β_I$ represents circular dichroism. If we suppose that $m_L=(n+β_R/2)+i(k+β_I/2)$ describes the optical characteristics of left-handed light, then $m_R=(n-β_R/2)+i(k-β_I/2)$ describes right-hand circularly polarized light. In our study we suppose that the particles contain chlorophyll *a* in concentration typical for cyanobacteria [11, 12]. Corresponding optical constants are specified in Nagdimunov et al. [12]. In the range of wavelengths used in the lab measurements [6, 7], the real index of refraction is 1.52 and a typical absorption is of order $10^{-2}$ to $10^{-3}$. Circular birefringence is of approximate order $10^{-8}$ to $10^{-9}$ and dichroism is slightly larger with an order of $10^{-7}$ to $10^{-8}$.

3. **Modeling Biological Objects as Aggregates**

To model biological objects, we used aggregates composed of multiple particles (monomers) due to the prevalence of aggregate structures in nature, e.g. chlorophyll in plant leaves, colonies of bacteria, protein aggregation, dust and some aerosols.

Two standard aggregation-accretion algorithms [12, 13] were used to generate the aggregates for our simulations: the ballistic particle-cluster aggregate (BPCA) model and the ballistic cluster-cluster aggregate (BCCA); samples of such aggregates are shown in inserts to Fig. 6. These two models are characterized by different porosity: aggregates created using the BPCA model have a porosity that is typically between 0.82 and 0.88, and aggregates created from the BCCA model have typical porosities ranging from 0.92 to 0.999, where the porosity is defined as a ratio of the volume of the voids in the aggregate to the total volume of the aggregate.

As a single generated aggregate is most commonly asymmetric, it will circularly polarize scattered light even if the material is completely achiral; this is another way of producing circularly polarized light [14, 15]. However, in real world situations, the multitude of aggregates with different structures would cancel each other's circular polarization if chiral materials are not involved. In the case of homochiral materials, considering one specific aggregate would affect the value of circular polarization as the effects of homochirality and aggregate's asymmetry are combined. In our computer simulations we corrected for this effect by calculating the polarization of two aggregates: the original aggregate and its mirror image. We then added the resultant Stokes parameters, producing the circular polarization that comes only from the difference in the indices of refraction, through which homochirality is reflected in the simulation. When we set optical rotation and circular dichroism to zero, the resultant circular polarization became zero as expected (see [16]).

In order to calculate the circular polarization of light that has been scattered by intrinsically asymmetric (homochiral) particles, we utilized Multiple Sphere T-matrix (MSTM) code for parallel computer clusters. This T-matrix code for simulating light scattering by a cluster of homogeneous spheres was introduced by Mackowski & Mishchenko (1996) [17], and modifications to support optically active materials are described in Mackowski et al. (2011) [16]. It is freely available at http://eng.auburn.edu/users/dmckwski/scatcodes/. Under this study, modifications of MSTM were performed to compute aggregate and its mirror image during one computer run that preserves the precision of each intermediate result without necessitating a large output.

4. **Results of Computer Simulations**

Light scattering computations were done for the wavelength range of 640 nm to 690 nm, a region where chlorophyll *a* has a large absorption band. We initially compared circular polarization that resulted from light scattering by aggregates, composed of chiral chlorophyll *a* monomers, to the lab measurements of cyanobacteria and macroscopic vegetation. Comparing Fig. 1 and Fig. 2 one can see that, the computational result and the lab measurements show very similar trends.

The most interesting feature of the laboratory and computer data is a change in sign of circular polarization that happens at the wavelength near largest absorption. This phenomenon is associated with Cotton effect [18, 19], where the circular dichroism and/or optical rotatory dispersion change sign through an electronic absorption band [7, 18].

Although the simulated trend of circular polarization spectra is well matched to the laboratory data and suggests that our model correctly reproduces polarization trends, the degree of circular polarization in the modeled results is several orders smaller than measured in the lab. To identify what determines the degree of circular polarization, and, thus, to establish the reason why computer modeling does not provide the values of circular polarization measured in the lab, we accomplished a computational survey for aggregates of different structure, size and number of monomers.

The results for aggregates of different number of monomers are shown in Fig. 3. Unfortunately, we were limited by computer resources and could study aggregates not larger than of 1024 monomers. Our simulations clearly show that increasing the number of monomers in the aggregate resulted in increased circular polarization. Other than increased degree, no significant change in the polarization trend as a function of scattering angle is observed.

Circular polarization also increases with the size of the individual monomers (Fig. 4), again preserving the overall polarization trend well. This is typical for both the spectral behavior (Fig. 4a) and for most scattering angles (Fig. 4b). The plots for monomers of radius ≥ 0.7 micron show an oscillatory behavior in circular polarization. Such a behavior is typical for individual monomers of the size comparable with wavelength. Below we discuss this case in more detail.

We also considered two aggregates of different porosity, BPCA of porosity 0.86 and BCCA of porosity 0.96, containing the same amount of identically sized monomers (Fig. 5). The less porous case displays increased degree of circular polarization across all wavelengths and scattering angles as compared to the more porous case. This is consistent with our earlier results [20], where we showed the same effect using extremes of porosity: a cube made of identical spheres, porosity 0.73, and a 3-D cross with porosity greater than 0.99. The paper reported that the degree of polarization increases significantly with the number of monomers for the cube but changed almost negligibly for the porous cross.

Notice also that despite a significant difference in the properties of aggregates, the behavior of circular polarization with scattering angle is almost identical for all cases, reaching a maximum at scattering angle 40-140° depending on the size of monomers. In the case of identical monomers, the polarization trend is unaffected by the number of monomers or porosity of the aggregate, these factors only influence the change in degree.

5. Discussion

In this section we discuss the physics behind the results of our computations, with the purpose of understanding what affects the formation of circular polarization in light scattering by optically active aggregates.

When considering aggregates, two competing mechanisms define the light-scattering properties: characteristics of individual monomers and electromagnetic interaction between them. Electromagnetic interaction refers to a situation where each monomer scatters light in a way that depends on the scattering of all other monomers. In other words, the exciting electromagnetic wave at each monomer can have a significant component due to the scattering of electromagnetic waves from neighboring monomers. This phenomenon is similar to multiple scattering, although the latter does not include near-field and coupling interactions and its mutual electromagnetic excitations occur simultaneously instead of being temporally discrete and ordered events. Recent papers by Kolokolova & Kimura [21] and Kolokolova & Mackowski [22] discussed electromagnetic interaction in aggregates and demonstrated that its strength depends on the number of monomers covered by an electromagnetic wave in a single period (on the light-path equal to one wavelength). One outcome of this effect is that compact aggregates are more strongly affected by electromagnetic interaction. Our results from Fig. 5 clearly demonstrate this, displaying a larger circular polarization for the more compact aggregate that is in all other characteristics (composition, size and number of monomers) identical to the more porous one.

The situation is more complicated when we consider aggregates containing differing numbers of monomers (Fig. 3). For aggregates with additional monomers, we can still partially attribute the increase in circular polarization to a larger contribution of interaction between the monomers. However, the increase can also be attributed in part to the added optically active material that light has to interact with.

Even more complicated is a situation having aggregates built of an equal number of monomers but with a different size of monomers for each aggregate (monomers in an individual aggregate are always of the same size) (Fig. 4). This situation combines electromagnetic interaction with characteristics of individual monomers that differ in both the larger amount of optically active material and the larger cross-section of bigger monomers. An interesting example of the influence of individual monomers is shown in Fig. 6a, where we present the results for two aggregates with an equal amount of optically active material, but in one case the material is stored in 32 monomers of radius 0.4 micron and in the other case the aggregate contains 256 monomers of radius 0.2 micron. To make them most compatible, the aggregates are of the same porosity, resulting in an equal characteristic radius for both. This case is thus also interesting as a comparison of two optically active aggregates of equal size. One would expect that either the results show circular polarization of the same degree (indicating that the amount of optically active material is the defining factor) or show that circular polarization of the aggregate with the larger number of monomers is greater (indicating strong electromagnetic interaction). However, Fig. 6a clearly demonstrates a larger value of circular polarization for the aggregate built of smaller number but bigger monomers. The most likely explanation of this fact is that the properties of individual monomers, specifically their cross-section, is the factor that dominates light-scattering in this situation. Fig. 6b shows that an individual sphere of radius 0.4 micron does produce a larger circular polarization than a sphere of radius 0.2 micron. Most likely, this original difference in the circular polarization of the individual monomers is responsible for the difference we see in aggregates built from these monomers.

The domination of properties of individual monomers is also evident in Fig. 3, where we see oscillating behavior of circular polarization for an aggregate with monomers of radius 0.7 micron, a size comparable to the wavelength. This provides further evidence that when a single wavelength covers one or just few monomers, electromagnetic interaction is weak and the characteristics of a single monomer are predominantly responsible for the resultant polarization. A similar effect was seen by Shen et al. [23], who noticed an oscillating behavior of linear polarization that disappeared when monomers of nonspherical shape or aggregates with a larger number of monomers were considered.

Thus we find that the formation of circular polarization is affected by both characteristics of individual monomers and electromagnetic interaction and that these effects are rather hard to disentangle from each other. Nevertheless, it is clear that electromagnetic interaction facilitates the increase of circular polarization. Consideration of larger and more compact aggregates would thus aid in the quantitative reproduction of the lab measurements [6, 7]; adjusting the size of individual monomers may also be of aid.

It is, however, unlikely that solely increasing the size of a single aggregate, its compactness or constituent monomer sizes would allow the simulations to reach measured values. In the measurements conducted by Sparks et al. [6, 7], samples presented optically thick media where circular polarization of a single aggregate was enhanced by multiple scattering between the aggregates (see similar effects in [24]). Combining T-matrix computations for optically active aggregates with radiative transfer techniques, to account for multiple scattering between the aggregates, should be a good addition to simulating light scattering by biological particles.

6. Conclusions

In contrast with minerals, rocks and other non-biological materials, biological particles produce non-zero circular polarization. This is a result of optical activity of biological materials that in turn results from homochirality of biological molecules. Recent measurements have also found that circular polarization undergoes a rapid change in value and potentially a change in sign inside some absorption bands for biological materials [6, 7].

Although we primarily investigated photosynthetic pigments, we expect that a similar rapid change in value and sign is present for certain chiral amino acids as multiple amino acids have been found to show optical characteristics akin to chlorophyll in both circular birefringence and dichroic spectra [8-10].

If the biological material forms an aggregated structure then the size of the monomers, their number in aggregate, and aggregate's porosity play a role in determining the resultant circular polarization. Our modeling showed that increasing the number of monomers in aggregate results in increased circular polarization. The increase is highly dependent on the porosity of aggregates and is larger for more compact aggregates; if the aggregate is extremely porous (~0.99 or more), then increase in circular polarization became negligible when the number of monomers was increased. This demonstrates that electromagnetic interaction between the monomers plays a significant role in the formation of circular polarization. However, the degree of circular polarization also increases with the size of the individual monomers, regardless of the porosity of the initial aggregate. Although we were not able to clearly disentangle to what degree electromagnetic interaction and to what degree the properties of individual monomers are responsible for the increase in circular polarization, we found that an increase in monomer size generally results in larger amounts of polarization than an increase in the number of monomers, at least for aggregates that contain less than 1024 monomers.

The spectral pattern of circular polarization found inside absorption bands of biological particles may allow not only for the remote identification of such particles, but the dependence of this pattern on the parameters of the aggregated structure could allow for some characterization of the detected materials.

Our study of biological particles is especially applicable to detection of biological or pre-biological materials in space; in particular, it allows us to make suggestions for future observations that target search of life in Solar System and on exoplanets. If attempting to detect signs of photosynthetic materials, the absorption bands that have the strongest polarization signatures are, most likely, associated with the wavelengths of maximum emission for the central star (or the Sun, if using this technique for solar system objects). Thus, the spectral range of maximum emission can be recommended for circular polarization measurements on planets both in our solar system and beyond. In the case of a search for homochiral amino acids (e. g. those found in meteorites, see [25, 26]), the wavelengths around 200 nanometers may be interesting because spectral behavior of the optical activity of amino acids bears a strong resemblance to chlorophyll's near this wavelength [8-10]. We also found that circular polarization typically reaches its peak at scattering angles of 40-140 degrees, and recommend observing in that range for maximum probability of detection.

Acknowledgment. This work has been supported by grant #NNX09AM97G from the NASA Astrobiology program.

**Figure Captions**

Fig. 1. Laboratory measurement of circular polarization spectra for cyanobacteria WH8101 shown at transmission (upper curve). Absorption shown in arbitrary units (lower curve). Adapted from [6].

Fig. 2. Simulated spectral dependence of circular polarization shown at transmission (upper curve). Absorption shown in arbitrary units (lower curve). The aggregate contains 256 monomers of radius 0.1 microns.

Fig. 3. Dependence of circular polarization on the number of monomers as a function of (a) wavelength and (b) scattering angle. The number of monomers, each with radius 0.1 microns, is specified in the legend. Spectral dependence, (a), is shown at scattering angle 90°. Wavelength for (b) is 660 nm.

Fig. 4. Dependence of circular polarization on the size of the monomers as a function of (a) wavelength and (b) scattering angle. Each aggregate contains 32 monomers; radii are given in microns at the top of the figure. Spectral dependence, (a), is shown at scattering angle 90° for the range of wavelengths 660 - 670 nm. Wavelength for (b) is 660 nm.

Fig. 5. Dependence of circular polarization on the porosity presented for BPCA and BCCA aggregates: (a) spectral dependence shown at transmission, and (b) scattering angle dependence at the wavelength 660nm. Each aggregate contains 32 monomers of radius 0.1 micron; porosity is 0.86 for BPCA and 0.95 for BCCA.  Example BPCA (left) and BCCA (right) aggregates are shown in the inset.

Fig. 6. Circular polarization vs. scattering angle (a) for two aggregates that contain an equal amount of optically active material, and (b) for the individual monomers of those aggregates. Wavelength is 660 nm.

Fig. 1

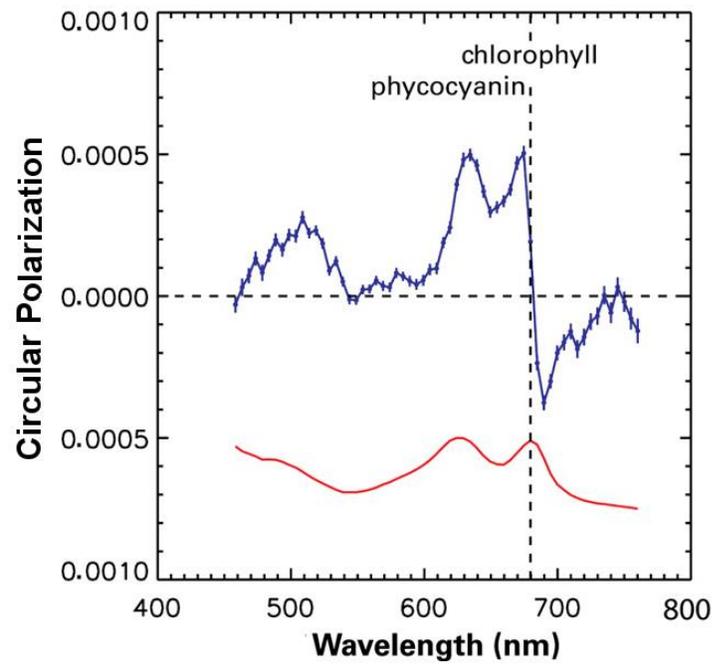

Fig. 2

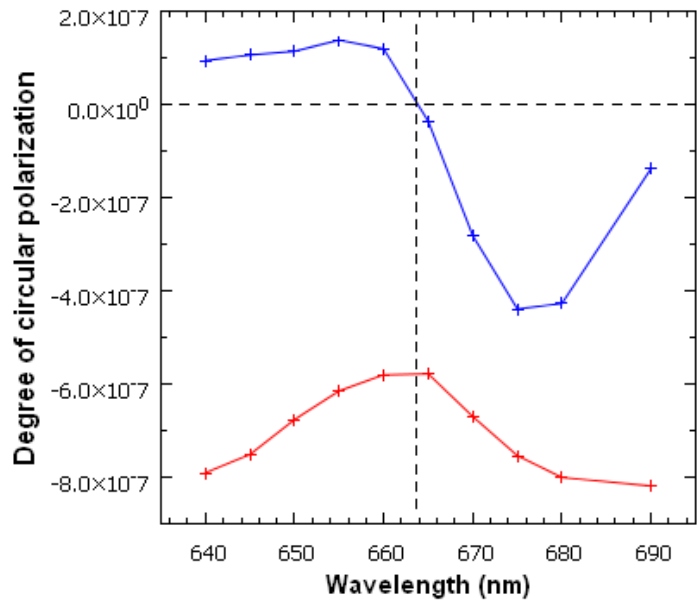

Fig. 3a

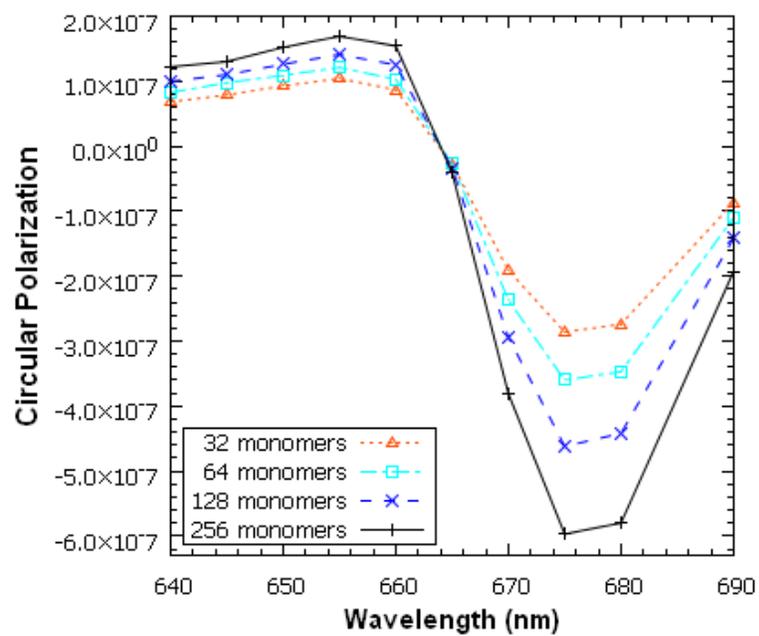

Fig. 3b

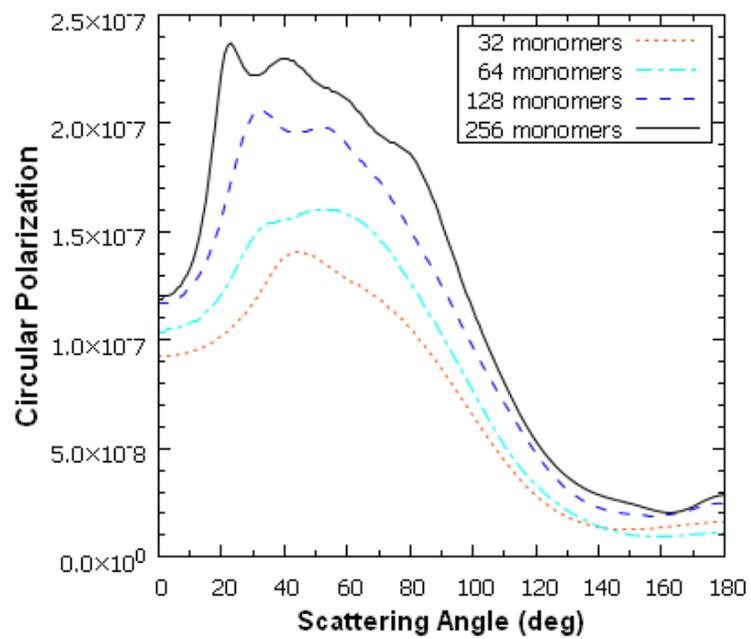

Fig. 4a

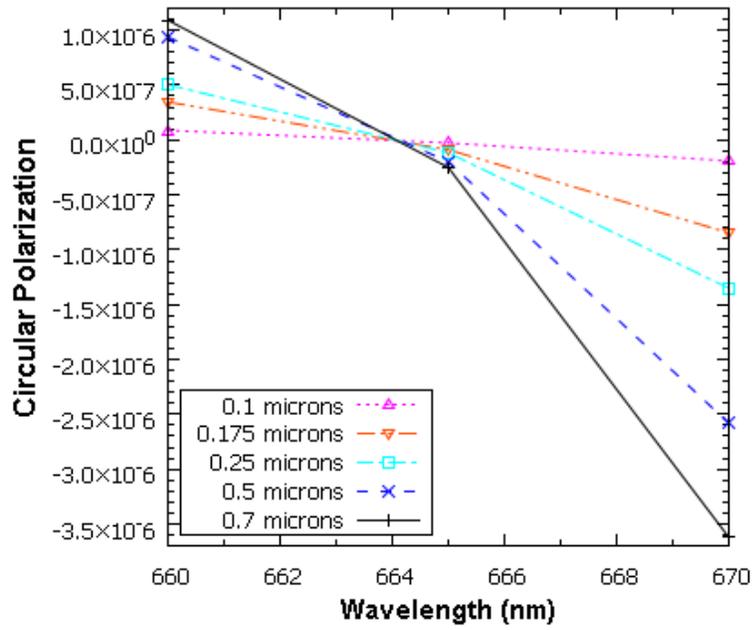

Fig. 4b

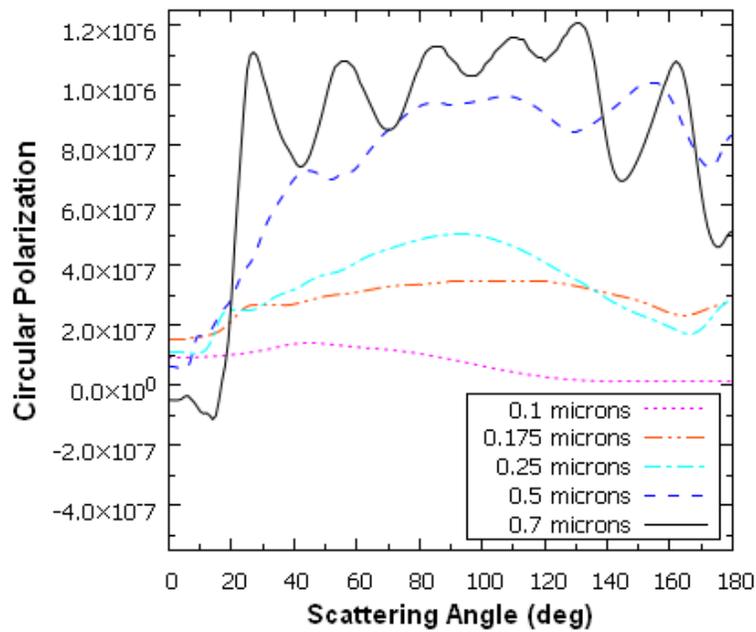

Fig. 5a

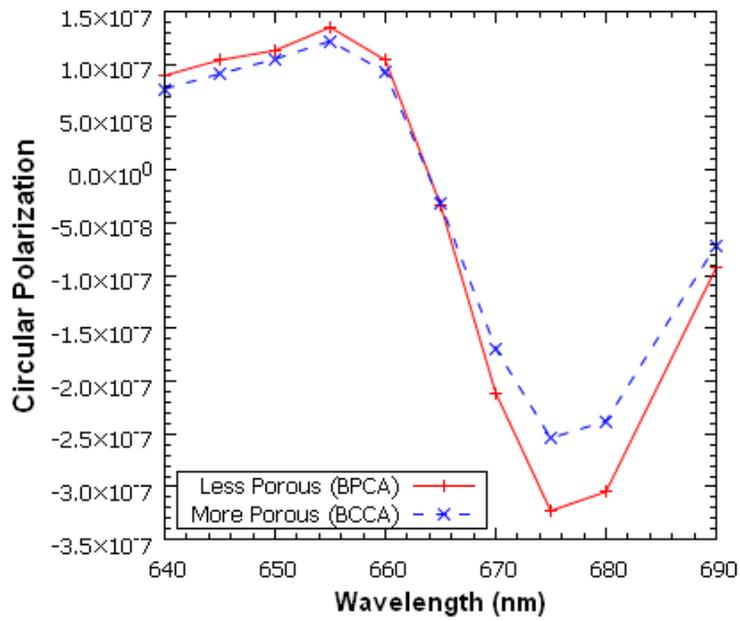

Fig. 5b

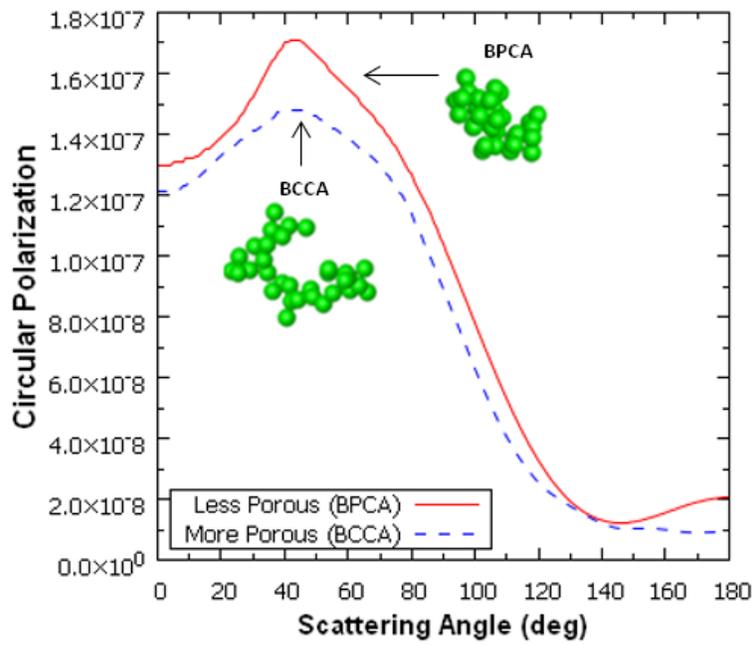

Fig. 6a

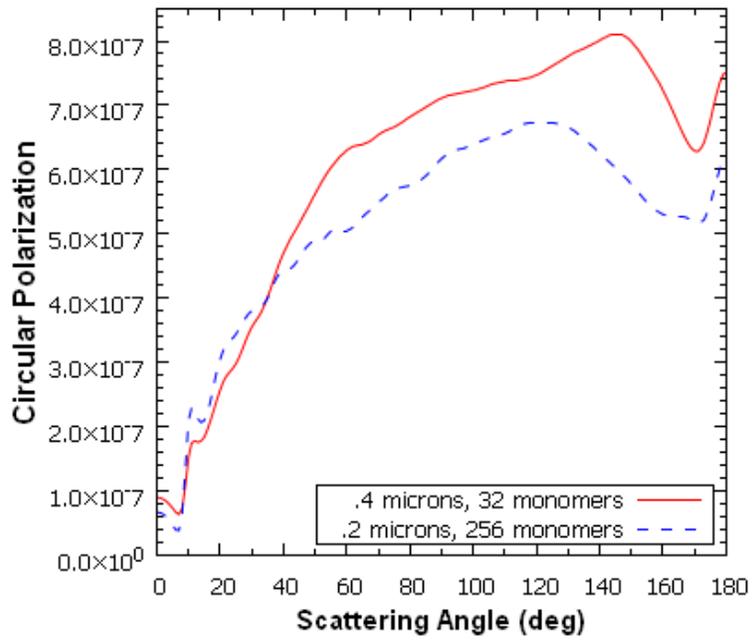

Fig. 6b

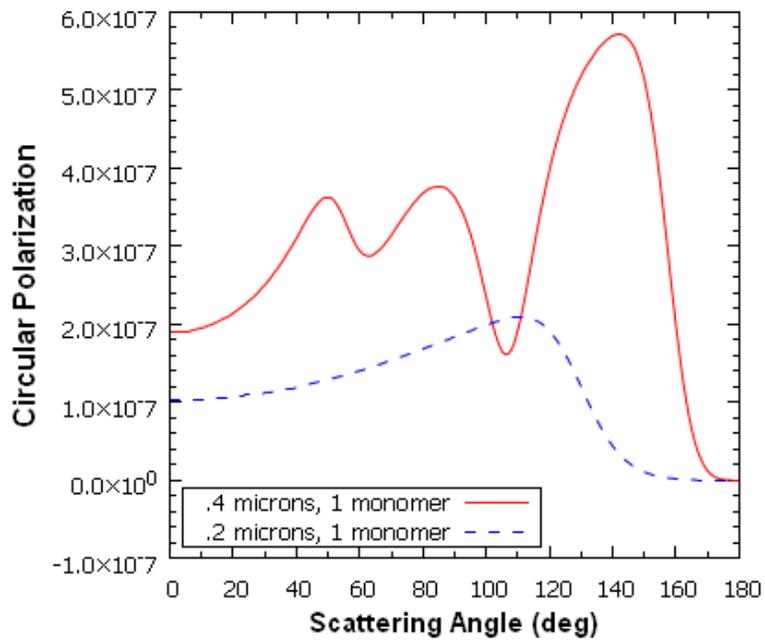